\documentclass[preprint,flushrt]{aastex}
\newcommand{\beq}{\begin{equation}}
\newcommand{\eeq}{\end{equation}}
\newcommand{\bea}{\begin{eqnarray}}
\newcommand{\eea}{\end{eqnarray}}
\newcommand{\bfig}{\begin{minipage}{3.3in}\bigskip}
\newcommand{\efig}{\bigskip\end{minipage}}
\newcommand{\mdot}{{\dot m}}
\newcommand{\rem}[1]{ }
\bibpunct[,]{(}{)}{;}{a}{}{,}
\begin{document}

\title{Self-Similar Hot Accretion Flow onto a Neutron Star}

\author{Mikhail V. Medvedev{}\altaffilmark{1} and Ramesh Narayan }

\affil{Harvard-Smithsonian Center for Astrophysics, 60 Garden Street, 
Cambridge, MA 02138 }
\altaffiltext{1}{Presently at the Canadian Institute for Theoretical
Astrophysics, University of Toronto, 60 St. George Street, Toronto,
ON, M6P 4B1, Canada; medvedev@cita.utoronto.ca; 
http://www.cita.utoronto.ca/$\sim$medvedev;
Also at the Institute for Nuclear Fusion, RRC ``Kurchatov
Institute'', Moscow 123182, Russia}
 
\begin{abstract}
We consider hot, two-temperature, viscous accretion onto a rotating,
unmagnetized neutron star.  We assume Coulomb coupling between
the protons and electrons, and free-free cooling from the electrons.
We show that the accretion flow has an extended settling region which
can be described by means of two analytical self-similar solutions: a
two-temperature solution which is valid in an inner zone,
$r\lesssim10^{2.5}$, where $r$ is the radius in Schwarzchild units; and
a one-temperature solution which is valid in an outer zone,
$r\gtrsim10^{2.5}$.  In both zones the density varies as $\rho\propto
r^{-2}$ and the angular velocity as $\Omega\propto r^{-3/2}$.  We 
solve the flow equations numerically and confirm that the analytical
solutions are accurate.

Except for the radial velocity, all gas properties in the self-similar
settling zone, such as density, angular velocity, temperature,
luminosity, angular momentum flux, are independent of the mass
accretion rate; these quantities do depend sensitively on the spin of
the neutron star.  The angular momentum flux is outward under most
conditions; therefore, the central star is nearly always spun-down.
The luminosity of the settling zone arises from the rotational energy
that is released as the star is braked by viscosity, and the
contribution from gravity is small; hence the radiative efficiency,
$\eta=L_{acc}/\dot Mc^2$, is arbitrarily large at low $\dot M$.  For
reasonable values of the gas adiabatic index $\gamma$, the Bernoulli
parameter is negative; therefore, in the absence of dynamically
important magnetic fields, a strong outflow or wind is not expected.
The flow is convectively stable, but may be thermally unstable.  The
described solution is not advection-dominated; however, when the spin
of the star is small enough, it transforms smoothly to an
advection-dominated branch of solution.

\end{abstract}

\section{Introduction}

At mass accretion rates less than a few per cent of the Eddington
rate, black holes (BHs) are believed to accrete via a hot,
two-temperature, radiatively inefficient, quasi-spherical,
advection-dominated accretion flow, or ADAF 
\citep{I77,NY94,NY95a,NY95,Abramowicz+95,Chen+95}.
The physical properties of ADAFs around BHs have been investigated by a
number of authors, and detailed spectral models have been applied to
observations of BH candidates 
\citetext{see \citealp{NMQ98} and \citealp{Kato+98} for reviews}.

At low mass accretion rates, accretion onto a neutron star (NS) is
also expected to occur via a hot, two-temperature flow
\citep{NY95,Yi+96}, but the properties of such flows have not been
investigated.  NS flows are expected to differ from BH ADAFs in
several respects.  (i) Whereas in the case of a BH the accreting
material flows freely and supersonically through the absorbing
boundary at the event horizon, in the case of a NS the radial velocity
of the material must decelerate to zero.  (ii) The accreting material
is expected to apply a spin-up or spin-down torque on the
NS. \citet{PN91} and \citet{Paczynski91} investigated the nature of
the torque for a cold thin disk, but the case of a hot flow has not
been studied.  (iii) For similar mass accretion rates (in Eddington
units), the luminosity of a NS accreting via an ADAF is likely to be
much higher than that of a BH because a NS has a surface while a BH
has an event horizon \citep{NY95,NGMc97,Menou+99}.  (iv) The spectra
are expected to be different.

We discuss in this paper the structure of a hot accretion flow around
a NS, or any other relativistic star with a surface.  The flow under
consideration is global and extends radially to very large distances
(at least thousands of NS radii or more) where it matches onto
appropriate outer boundary conditions.  We do not attempt a detailed
analysis of the boundary layer region near the NS, where the accretion
flow meets the star.  We present only an approximate analysis of this
region, which extends at most a few NS radii above the stellar surface
\citetext{for a more detailed discussion of the physics of the
boundary layer see \citealp{PN91,Paczynski91, TLM98,TO99}}.

The paper is organized as follows. We show in \S\ref{S:SSS} that there
is a radially extended region around the NS where the flow ``settles''
with a radial velocity much less than the local free-fall velocity.
We obtain a self-similar solution for this settling region and show
that, surprisingly, the density and temperature of this zone are
independent of the mass accretion rate.  In \S\ref{S:PROP}, we discuss
physical properties of the accretion flow. In \S\ref{S:NUM} we compare
the analytical results with numerical computations and in
\S\ref{S:TRANS} we discuss the relationship between the settling flow
and an ADAF.  We conclude with a discussion in \S\ref{S:SandD}.

\section{Self-Similar Settling Solution }
\label{S:SSS} 

We consider a steady, rotating, axisymmetric, quasi-spherical,
two-temperature accretion flow onto a star with a surface.  We use the
height-integrated form of the viscous hydrodynamic equations 
\citep{I77,A+88,Paczynski91,NY94}:
\bea
\displaystyle &\displaystyle \dot M=4\pi R^2\rho v , & \label{mdot}\\
&\displaystyle v\frac{d v}{d R}=\left(\Omega^2-\Omega_K^2\right)R
-\frac{1}{\rho}\frac{d}{dR}\left(\rho c_s^2\right) , & \label{mom}\\
&\displaystyle 4\pi\alpha\frac{\rho c_s^2R^4}{\Omega_K}\frac{d\Omega}{dR} 
=\dot J-\dot M\Omega R^2 , & \label{omega}\\ 
&\displaystyle \rho v T_p\frac{d s_p}{dR}
=\frac{\rho vc^2}{(\gamma_p-1)}\frac{d\theta_p}{dR}
-vc^2\theta_p\frac{d\rho}{dR}=(1-\delta)q^+-q_{\rm Coul} , & \label{energy-p}\\
&\displaystyle \rho_e v T_e\frac{d s_e}{dR}
=\frac{\rho_e vc^2}{(\gamma_e-1)}\frac{d\theta_e}{dR}
-vc^2\theta_e\frac{d\rho_e}{dR}=\delta\,q^++q_{\rm Coul}-q^- ,& \label{energy-e}
\eea 
where $\dot M$ is the mass accretion rate, $R$ is the spherical
radius, $\rho$ is the mass density of the accreting gas, $v$ is the
radial infall velocity, $\Omega$ is the angular velocity,
$\Omega_K(R)=(GM/R^3)^{1/2}$ is the Keplerian angular velocity,
$c_s^2=c^2(\theta_p+\theta_e m_e/m_p)$ is the square of the isothermal
sound speed, $T_{p,e}$ are the temperatures of protons and electrons,
$\theta_{p,e}=k_BT_{p,e}/m_{p,e}c^2$ are the corresponding
dimensionless temperatures, $\alpha$ is the Shakura-Sunyaev viscosity
parameter, $\dot J$ is the rate of accretion of angular momentum,
$s_p$ and $s_e$ are the specific entropies of the proton and electron
fluids, $\rho_e\simeq(m_e/m_p)\rho$ is the mass density of the
electron fluid, $\gamma_p$ and $\gamma_e$ are the adiabatic indices of
protons and electrons (which, in general, may be functions of $T_p$
and $T_e$), and $q^+$, $q^-$, and $q_{\rm Coul}$ are the viscous
heating rate, radiative cooling rate, and energy transfer rate from
protons to electrons via Coulomb collisions, per unit mass. 
We have assumed that a fraction $\delta$ of the viscous heat goes into 
electrons and a fraction $1-\delta$ into protons; it is usually 
assumed in ADAF models that $\delta\ll1$, but our analysis is general 
for any value of $\delta$ between 0 and 1.  
Equations (\ref{mdot})--(\ref{energy-e}) describe the conservation of mass,
radial momentum, angular momentum, proton energy and electron energy,
respectively.

For simplicity, we have assumed in equation (\ref{mdot}) 
that the flow is spherical.  A more accurate treatment
would replace $R^2$ with $RH$, where the scale height
$H=c_s/\Omega_k$. This would introduce minor differences 
in some of the quantitative results.

In the case of accretion onto a NS we expect the flow to slow down as
it settles on the stellar surface, and we expect the density in this
settling zone to be significantly higher than for a BH.  The increased
density would cause more efficient transfer of energy from protons to
electrons via Coulomb collisions and more efficient radiation from the
electrons.  As we show below, this leads to a flow in which $q^+$,
$q^-$ and $q_{\rm Coul}$ are all of the same order, which is very
different from the case of a BH ADAF, where $q^+\gg q^-,\, q_{\rm
Coul}$.  Another feature of the settling zone, again the result of the
large density, is that optically thin bremsstrahlung cooling (which is
sensitive to $\rho$) dominates over self-absorbed synchrotron cooling.
We therefore neglect synchrotron emission in our analysis.  For
simplicity, we neglect also thermal conduction.  Comptonization is
important over part of the settling zone.  However, we have not been
able to derive useful analytical results with both bremsstrahlung and
Comptonization included.  Therefore, we neglect Comptonization in this
section, and discuss its effects in \S\ref{S:PROP-COMPT}.

The set of equations (\ref{mdot})--(\ref{energy-e}) must satisfy certain 
boundary conditions at the neutron star.  First, as the flow approaches the
surface of the star at $R=R_{NS}$, the radial velocity must become
very much smaller than the local free-fall velocity.  Second, the
angular velocity must approach the angular velocity of the star
$\Omega_{NS}$.  We use the dimensionless parameter
\beq
s\equiv {\Omega_{NS}\over\Omega_K(R_{NS})}
\label{def:s}
\eeq
to represent the spin of the NS.

The radius of the star, and its spin, are the two principal boundary
conditions applied at the inner edge of the accretion flow.  We assume
that the star is unmagnetized, so there are no magnetospheric effects
to consider.  Two outer boundary conditions, namely the temperature
and angular velocity of the gas, are determined by the properties of
the gas as it is introduced into the accretion flow on the outside
(e.g. from an ambient medium or from a different type of accretion
flow such as a thin disk).  These outer boundary conditions have
little effect on the interior of the flow \citetext{see
\S\ref{S:TRANS} of this paper, and \citealp{NKH97}; but see also
\citealp{Yuan99}}.  An additional important boundary condition is the
mass accretion rate $\dot M$, which is determined by external
conditions and which we take to be constant.

\subsection{Inner Settling Solution}
\label{S:SSS-I}

We consider first the inner region of the flow, $R_{NS}\la R\la
10^{2.5}R_S$, where $R_S=2GM/c^2$ is the Schwarzchild radius.  In this
region we expect a two-temperature plasma, with $T_p>T_e$, in which
the electrons are relativistic and the protons are non-relativistic:
$\theta_e\gg1,\ \theta_p\ll1$.  The viscous heating rate of the gas,
the energy transfer rate from the protons to the electrons via Coulomb
collisions, and the cooling rate of the electrons via bremsstrahlung
emission are given by
\bea 
q^+&=&\alpha\frac{\rho c^2_s
R^2}{\Omega_K}\left(\frac{d\Omega}{dR}\right)^2,
\label{q+}\\
q_{\rm Coul}&=&Q_{\rm Coul}\,\rho^2\frac{\theta_p}{\theta_e} ,\quad
Q_{\rm Coul}=4\pi r_e^2 \ln{\Lambda} \frac{m_ec^3}{m_p^2}, 
\label{qc}\\
q^-&=&Q_{\rm ff,R}\,\rho^2\theta_e ,\quad
Q_{\rm ff,R}=48\alpha_f r_e^2\frac{m_ec^3}{m_p^2} ,
\label{qff}
\eea 
where $\alpha_f$ is the fine structure constant, $r_e$ is the
classical electron radius, $\ln{\Lambda}\simeq20$ is the Coulomb
logarithm, $c_s^2\simeq c^2\theta_p$, and we have neglected logarithmic
corrections to the relativistic free-free emissivity. The subscript
``R'' in $Q_{\rm ff,R}$ denotes relativistic bremsstrahlung.

We now make a number of simplifications in equations
(\ref{mdot})--(\ref{energy-e}).  First, we neglect the radial velocity
term $vdv/dR$ in equation (2).  Second, we assume that $\dot J$
dominates over $\dot M\Omega R^2$ on the right-hand-side of equation
(\ref{omega}) and we neglect the latter term.  Third, we neglect the
entropy terms in equations (\ref{energy-p}), (\ref{energy-e}); that
is, we assume that the heating, cooling and energy transfer terms in
these equations dominate over the entropy gradient terms.  All of
these assumptions are justified {\it a posteriori} below. The
equations then read
\bea 
\displaystyle &\displaystyle \dot M=4\pi R^2\rho v , &
\label{mdot1}\\ &\displaystyle \left(\Omega^2-\Omega_K^2\right)R
=\frac{1}{\rho}\frac{d}{dR}\left(\rho c_s^2\right) , & \label{mom1}\\
&\displaystyle 4\pi\alpha\frac{\rho c_s^2
R^4}{\Omega_K}\frac{d\Omega}{dR} =\dot J , & \label{omega1}\\
&\displaystyle (1-\delta)q^+= q_{\rm Coul}=q^--q^+\delta, & \label{energy1}
\eea 
where $q^+,\ q_{\rm Coul}, \textrm{ and } q^-$ are given by
equations (\ref{q+})--(\ref{qff}).

It is straightforward to show that the simplified equations 
(\ref{mdot1})--(\ref{energy1}) have the following self-similar solution,
\bea
& \rho=\rho_0r^{-2},\quad \theta_p=\theta_{p0}r^{-1},\quad 
\theta_e=\theta_{e0}r^{-1/2}, & \nonumber\\
& \Omega=\Omega_0r^{-3/2},\quad v=v_0r^0, &
\label{sss}
\eea 
where $r=R/R_S$ is a dimensionless radius.  The normalization
coefficients are uniquely related to each other as follows
\begin{mathletters}
\bea
\theta_{p0}&=&\frac{1}{6}\left(1-s^2\right) , \\
\theta_{e0}&=&\left(\frac{Q_{\rm Coul}}{Q_{\rm ff,R}}
\frac{\theta_{p0}}{(1-\delta)}\right)^{1/2}
=\left(\frac{\pi\ln\Lambda}{12\alpha_f}
\frac{\theta_{p0}}{(1-\delta)}\right)^{1/2}
\simeq10.9\left(\frac{1-s^2}{1-\delta}\right)^{1/2},\\
\Omega_0&=&\Omega_{K0}\,s\simeq7.19\times10^4\,m^{-1}s\textrm{  rad/s} ,\\
\rho_0&=&\frac{9c^2\Omega_{k0}}{4Q_{\rm Coul}}\,\alpha(1-\delta)\theta_{e0}s^2 
\simeq8.10\times10^{-4}m^{-1}\alpha(1-\delta)\theta_{e0}s^2\textrm{  g/cm}^3 ,\\
v_0&=&\frac{\dot M}{4\pi R_S^2\rho_0} ,\\
\dot J&=&-6\pi\alpha c^2R_S^3\rho_0\theta_{p0}s
=-\frac{9\pi}{2\sqrt{3}}\frac{(GM)^2c}{\sqrt{Q_{\rm Coul}Q_{\rm ff,R}}}
\,\alpha^2(1-\delta)^{1/2}s^3\left(1-s^2\right)^{3/2} .
\eea
\label{norm}
\end{mathletters}
Here, $\Omega_{K0}=\Omega_K(R_S)$ and the dimensionless spin
parameter is $s=\Omega(R_S)/\Omega_{K0}=\Omega_{NS}/\Omega_K(R_{NS})
=\Omega(R)/\Omega_K(R)$, as introduced in equation (\ref{def:s}).  
Recall that $s$ is a boundary condition of the problem.  

The range of $r$ over which the solution is valid is determined by the
twin requirements that the protons be non-relativistic and that the
electrons be relativistic.  The former condition is satisfied for any
$r>1$, while the latter condition requires $r<\theta_{e0}^2\simeq120$.
A third condition is that Comptonization should be negligible (since
we have assumed this).  As we show in \S\ref{S:PROP-COMPT}, this last
condition requires $r>\textrm{few tens}$, with the exact limit
depending on the NS spin, $s$, and viscosity, $\alpha$.

The radial velocity of the solution is independent of $r$, whereas the
local free-fall velocity varies as $v_{ff}=c/\sqrt{2r}$.  Thus,
$v/v_{ff}$ decreases with decreasing radius.  This shows that the
solution corresponds to a settling flow and that it is quite different
from self-similar ADAFs around BHs, where $v/v_{ff}$ either is constant 
\citep{NY94,NY95a,Manmoto+00} or increases with decreasing $r$ 
\citep{NIA99,QG99}. Although $v/v_{ff}$ is quite small as the flow
approaches the NS surface, $v$ itself is still fairly large.  At the
NS surface, $v$ must reduce substantially from its self-similar value.
As the numerical results of \S\ref{S:NUM} show, this happens in a boundary 
layer where the accreting material cools catastrophically to a temperature
that is orders of magnitude below virial.  The boundary layer is
distinct from the settling zone which is described by the above
self-similar solution.

The angular velocity of the gas is a fixed fraction of the local
Keplerian angular velocity, the ratio being determined by the
dimensionless spin $s$ of the star.  The gas in the settling solution
radiates most of the energy dissipated through viscosity.  In fact,
the rates of viscous heating, Coulomb energy transfer and radiative
emission are all equal, which is achieved by a suitable choice of the
density, electron temperature and proton temperature in the gas.  The
two temperatures have universal forms, with only a weak dependence on
$s$, while the density has a strong dependence on $s$.

The most outstanding feature of the self-similar solution is that,
except for the radial velocity, none of the other gas parameters has
any dependence on $\dot M$.  

The fact that $\dot J$ is negative implies that the accretion flow
removes angular momentum from the star and spins it down.  This behavior is
quite different from that seen in thin disks 
\citep{PN91,Paczynski91}, where for most choices of the stellar spin
parameter $s$, the accretion disk spins up the star with a torque
$\dot J_{\rm thin}\approx\dot M\Omega_K(R_{NS})R_{NS}^2$.  Only when $s$ is
very close to unity does the torque become negative.
In contrast, for the self-similar solution derived here, the torque is
negative for all values of $s$ (except extremely small values, see the
discussion in \S\ref{S:NUM}).  Moreover, $\dot J$ is independent of $\dot M$.
Equivalently, the dimensionless torque, $j=\dot J/\dot
M\Omega_K(R_{NS})R_{NS}^2$, which is $\sim1$ under most 
conditions for a thin disk, here takes on the value 
\bea 
j&=&-\frac{\sqrt{\pi}}{8\sqrt{2}}\frac{m_p}{m_e}
\frac{\eta}{\sqrt{\alpha_f\ln\Lambda}}\,
r_{NS}^{-1/2}(1-\delta)^{1/2}\alpha^2\mdot^{-1}s^3\left(1-s^2\right)^{3/2}
\nonumber\\
&\simeq&-43{\alpha^2}{\mdot}^{-1}s^3 \left(1-s^2\right)^{3/2} ,
\label{j}
\eea 
where $\mdot=\dot M/\dot M_{\rm Edd}$ is the mass accretion rate in 
Eddington units, with $\dot M_{\rm Edd}=1.39\times10^{18}m\textrm{ g/s}$, 
(for a nominal $\eta=0.1$), $r_{NS}\approx3$.  
Note that $-j$ could be very large at low $\dot m$.

We now check under what conditions the approximations we made earlier
are valid.  First, we neglected the
term $vdv/dR$ in equation (\ref{mom}).  This is obviously valid since
the self-similar solution has $v= {\rm constant}$.

Second, we assumed that $|\dot J|\gg|\dot M\Omega R^2|$ in equation
(\ref{omega}). Using (\ref{j}) this condition may be cast into the
form $|j|\gg sr^{1/2}/\sqrt{3}$, or equivalently, 
\beq
\mdot\ll74\alpha^2s^2r^{-1/2}.
\label{constraint}
\eeq 
For $\alpha\sim0.1,\
s\sim0.3$, and assuming a radial extent of $r\sim10^2$ for the flow,
we require $\mdot\la7\times10^{-3}$. A
direct numerical simulation (\S\ref{S:NUM}) shows that the
analytical solution is valid even for values of $\dot m$ that are  a 
factor of a few larger than this limit.

Third, we neglected the entropy terms $\rho_{p,e}vT_{p,e}ds_{p,e}/dR$
in equations (\ref{energy-p}) and (\ref{energy-e}).  For the protons,
using the solution (\ref{sss}), we can show that the left-hand-side of
equation (\ref{energy-p}) varies as $r^{-4}$ while the
right-hand-side varies $r^{-4.5}$.  Thus, with decreasing $r$, the
entropy term becomes progressively less important than the other terms,
thereby confirming the validity of the
approximation.  In the case of the electrons, the entropy is always
small since $s_e\sim(m_eT_e/m_pT_p)s_p\ll s_p$.

\subsection{Outer Settling Solution}
\label{S:SSS-O}

For $r>10^{2.5}$, both protons and electrons are non-relativistic and
the solution described in the previous subsection is not valid.
Interestingly, another self-similar solution may be derived for this
region of the flow.  This solution has a nearly one-temperature
plasma with $T_p-T_e\ll T_p,\,T_e$.  
The free-free cooling takes the form
\beq
q^-\sim Q_{\rm ff,NR}\rho^2\theta_e^{1/2}, \quad 
Q_{\rm ff,NR}=5\sqrt{2}\pi^{-3/2}\alpha_f\sigma_T\frac{m_ec^3}{m_p^2},
\eqnum{\ref{qff}$'$} 
\eeq
where $\sigma_T$ is the Thompson cross-section, and the subscript
${\rm NR}$ stands for non-relativistic.  Since the gas is effectively
one-temperature, Equation (\ref{energy1}) simplifies to
\beq 
q^+\simeq q^- .  \eqnum{\ref{energy1}$'$}
\eeq 
We do not need to consider the Coulomb transfer rate $q_{\rm Coul}$,
since this quantity is proportional to $(T_p-T_e)$ and
can be adjusted to have the right magnitude with small adjustments
of the two temperatures.  In the non-relativistic regime, $\theta_e<1,\ 
\theta_p\sim (m_e/m_p)\theta_e\ll1$, and the Coulomb transfer rate is
\beq
q_{\rm Coul}=\frac{3}{\sqrt{2\pi}}\frac{m_e}{m_p}\frac{\sigma_Tc}{m_p^2}
\ln\Lambda\,\rho^2\frac{kT_p-kT_e}{\theta_e^{3/2}e^{-1/\theta_e}}.
\eeq
>From the condition $q_{\rm Coul}\simeq q^-$ it follows that
\beq
kT_p-kT_e\simeq\frac{10}{3\pi^2}\frac{\alpha_f}{\ln\Lambda}\sqrt{m_pm_e}c^2
\theta_e^2e^{-1/\theta_e},
\eeq
which is exponentially small for $\theta_e<1$.

The flow is again described by the self-similar solution (\ref{sss}),
with the following two exceptions:
\beq 
\theta_e=\theta_{e0}r^{-1}, \qquad
\theta_{e0}=\frac{m_p}{m_e}\,\theta_{p0} ,
\eqnum{\ref{norm}b$'$} \eeq 
\beq 
\rho=\rho_0r^{-2}, \qquad 
\rho_0 =\frac{9\alpha c^2\Omega_{K0}}{2Q_{\rm ff,NR}}\, 
\frac{\theta_{p0}s^2}{\theta_{e0}^{1/2}}
=0.12m^{-1}\theta_{p0}^{1/2}\alpha s^2\textrm{  g/cm}^3,
\eqnum{\ref{norm}d$'$} 
\eeq
where we have used the fact that the total pressure $p=p_p+p_e=2p_p$.

This solution is valid only if $\dot m$ is quite small,
cf. equation (\ref{constraint}):
\beq
\mdot<2.2\times10^{-3}\alpha_{0.1}^2s_{0.3}^2r_3^{-1/2},
\label{out-constr}
\eeq 
where $r_3=r/10^3$, $\alpha_{0.3}=\alpha/0.3$, and $s_{0.1}=s/0.1$. 
For greater $\mdot$, a self-similar power-law 
solution does not exist; numerically computed solutions exhibit 
non-power-law behavior, as discussed in \S\ref{S:NUM}.

\section{Properties of the self-similar solution}
\label{S:PROP} 

\subsection{Spin-Up/Spin-Down of the Neutron Star}
\label{S:PROP-SPIN}

The rate of spin-up of the accreting NS is given by
\bea
\frac{d}{dt}\left(I_{NS}\Omega\right)&=&\dot J-\dot M\Omega(R_{NS})R_{NS}^2 
\nonumber\\
&\simeq&-43s^3\alpha^2\dot M_{\rm Edd}\Omega_K(R_{NS})R_{NS}^2,
\label{spin-up}
\eea
where $I_{NS}$ is the moment of inertia of the NS.  We have made use of
the fact that $|\dot J|\gg|\dot M\Omega(R_{NS})R_{NS}^2|$ for the 
self-similar solution, and used equation (\ref{j}) for $\dot J$.
The negative sign in the final expression implies that the accretion
flow spins down the star. The above equation is for an unmagnetized NS. 
If the NS has a magnetosphere, the inner edge of the accretion flow is 
at the magnetospheric radius, $R_m$. In this case, let us define $s$ by 
$\Omega_{NS}=s\Omega_K(R_m)=s\Omega_K(R_{NS})(R_m/R_{NS})^{-3/2}$. 
Substituting this in equation (\ref{spin-up}) with $I_{NS}=constant$ 
and integrating, we obtain 
\beq
s=\frac{s_0}{\sqrt{1+t/\tau}}, \qquad
\tau=\frac{I_{NS}}{86s_0^2\alpha^2\dot M_{\rm Edd}R_{NS}^2}
\left(\frac{R_m}{R_{NS}}\right)^{-3/2},
\eeq
where ${s_0=s({t=0})}$. The same result is valid for an unmagnetized NS
by setting $R_m=R_{NS}$.
The quantity $\tau$ is the characteristic spin-down time of the NS. 
For a spherical NS of constant density,
$I_{NS}=2MR^2_{NS}/5=(0.8\times10^{33}\textrm{ g})mR_{NS}^2$.
Substituting this expression, we obtain the spin-down rate
$\dot P_{NS}/P_{NS}=\tau^{-1}$ with 
\beq
\tau\simeq6.7\times10^{12}s^{-2}\alpha^{-2}
\left(\frac{R_m}{R_{NS}}\right)^{-3/2}\textrm{ s}
=2\times10^{8}s_{0.1}^{-2}\alpha_{0.1}^{-2}
\left(\frac{R_m}{R_{NS}}\right)^{-3/2}\textrm{ yr} .
\eeq 
Note the remarkable fact that the spin-down time scale is independent
of the mass of the NS, and the mass accretion rate! For the magnetic
case, the rate depends on the radius ratio ${R_m}/{R_{NS}}$.

It is customary to express the spin-down rate as ${\dot P_{NS}}/{P_{NS}^2}$.
Writing 
\beq
P_{NS}=\frac{2\pi}{s\Omega_K(R_{NS})}\left(\frac{R_m}{R_{NS}}\right)^{3/2}
\eeq
and $\Omega_K(R_{NS})\simeq10^4m_{1.4}^{-1}\textrm{ rad/s}$, where 
$R_{NS}=3R_S$ and $m_{1.4}=M/(1.4M_{\sun})$, we obtain
\beq
\frac{\dot P_{NS}}{P_{NS}^2}
\simeq2.4\times10^{-10}m_{1.4}^{-1}\alpha^2s^3\textrm{ s}^{-2}
=2.7\times10^{-12}m_{1.4}^{-1}\alpha_{0.3}^2s_{0.5}^3\textrm{ s}^{-2},
\eeq
where $s_{0.5}=s/0.5$.
This spin-down rate is in good agreement with observational data
on the spin-down of X-ray pulsars for which \citet{YWV97} invoked ADAFs: 
4U~1626-67 has $\dot P/P^2\approx8\times10^{-13}\textrm{ s}^{-2}$ 
and $P=7.7\textrm{ s}$; OAO~1657-415 has 
$\dot P/P^2\approx2\times10^{-12}\textrm{ s}^{-2}$ and $P=38\textrm{ s}$,
and GX~1+4 has $\dot P/P^2\approx3.7\times10^{-12}\textrm{ s}^{-2}$ and 
$P=122\textrm{ s}$. Since the spin-down rate is quite sensitive to 
$\alpha$ and $s$, the observed data in individual systems can be fitted
by small adjustment of these parameters.

\subsection{Luminosity}
\label{S:PROP-LUMIN}

In computing the luminosity of the accretion flow, we must allow
for the energy release in both the boundary layer and the self-similar 
settling zone. We calculate their luminosities separately.

Radiation from the self-similar settling flow may be calculated
following the methods described by \citet{PN95} for a thin disk.
This method assumes that the luminosity at a given radius is determined 
by the local viscous energy production. This is a legitimate 
approximation for the settling flow in which $q^-=q^+$.
Keeping only the dominant terms, we find
\beq
L_{SS}=\frac{GM_{NS}\dot M}{R_{in}}\left(1+\frac{1}{2}s^2-js\right)
+\dot M\int_{P_{\rm in}}^{P_{\rm out}}\frac{dP}{\rho},
\eeq
where $R_{in}=R_{NS}+\Delta_{BL}$ is the inner radius of the self-similar
zone and $\Delta_{BL}\ll R_{NS}$ is the thickness of the boundary layer.
Here the first two terms, $(1+s^2/2)$, represent the luminosity,
associated with potential energy of the infalling gas,  
$L_{\rm pot}$, the third term $-js$ is the luminosity, associated with 
the rotational energy extracted from the star, $L_{\rm rot}$ 
(note, $j<0$ in the self-similar solution), and the final integral 
is the ``enthalpy correction'', $L_{\rm enth}$. Using the analytical 
solution (\ref{sss})--(\ref{norm}) and assuming ${P_{\rm out}}=0$ for 
simplicity, we obtain
\begin{mathletters}
\bea
L_{\rm pot}&=&\dot M_{\rm Edd}c^2\frac{\mdot}{2r_{NS}}
\left(1+\frac{s^2}{2}\right),\\
L_{\rm rot}&=&43\dot M_{\rm Edd}c^2\frac{\alpha^2}{2r_{NS}}
(1-\delta)^{1/2}s^4\left(1-s^2\right),\\
L_{\rm enth}&=&-\dot M_{\rm Edd}c^2\frac{\mdot}{2r_{NS}}\left(1-s^2\right) .
\eea
\end{mathletters}
Note that the leading terms in $L_{\rm pot}$ and $L_{\rm enth}$ 
cancel each other exactly. The luminosity of the self-similar settling zone 
is thus
\beq
L_{SS}\simeq
6.2\times10^{34}mr_3^{-1}\mdot_{-2}s_{0.1}^2
+8.9\times10^{33}mr_3^{-1}\alpha_{0.1}^2s_{0.1}^4 
\textrm{ erg\,s}^{-1},
\label{L-settl}
\eeq 
where $\mdot_{-2}=\mdot/0.01$, $s_{0.1}=s/0.1$, $r_3=r_{NS}/3$,
and we have assumed $s\ll1$.  Note that luminosity not associated with
with dissipation of rotational energy, represented by the first term
in equation (\ref{L-settl}), is much less than the commonly assumed
$\sim GM_{NS}\dot M/R_{NS}$. This is because the negative enthalpy
term has large magnitude, as a result of the fact that the settling
flow is akin to a pressure supported, quasi-stationary atmosphere.

The second term in equation (\ref{L-settl}) is the luminosity of the 
settling zone.  Since the self-similar solution for this zone is 
independent of $\dot m$, the luminosity too shows no $\dot m$ dependence.  
Indeed, the luminosity remains finite even as $\dot m\to0$.  How is this
possible, and where does the energy come from?  The answer is that the
luminosity of the settling zone is supplied by the central star.  As
the star spins down, it does work on the accretion flow and the
energy released comes out as bremsstrahlung radiation.

The boundary layer luminosity requires a different method of
calculation since viscous energy production is negligible in this
zone: $\Omega\simeq$constant, and so
$q^+\propto\left(d\Omega/dR\right)^2\simeq0$.  As the accreting gas
cools in the boundary layer, starting from a nearly virial temperature
$\sim10^{12}$~K on the outside down to the NS temperature $\sim10^7$~K
near the surface, the thermal energy in the gas is emitted as
radiation.  To estimate the luminosity, we use the energy balance
equation, which is the sum of equations
(\ref{energy-p}),(\ref{energy-e}): 
\beq 
-q^-=\frac{\rho v}{\gamma-1}\frac{dc_s^2}{dR}-vc_s^2\frac{d\rho}{dR}
=\frac{\gamma}{\gamma-1}\rho v\frac{dc_s^2}{dR}-v\frac{dP}{dR}.  
\eeq
We can neglect the $dP/dR$ term because the pressure $P$ is essentially
constant in the boundary layer. To obtain the luminosity we integrate over
the boundary layer 
\beq 
L_{BL}=\int q^- 4\pi R^2 dR
=-\int\frac{\gamma}{\gamma-1}4\pi R^2\rho v\frac{dc_s^2}{dR}\, dR
=\frac{\gamma}{\gamma-1}\dot M\Delta c_s^2.  
\eeq 
Since $c_s^2$ starts from nearly virial value and reaches close to zero, 
$\Delta c_s^2\simeq GM_{NS}/R_{NS}$. More precisely, 
$\Delta c_s^2=c^2\Delta(\theta_p+\theta_e)\simeq c^2\theta_{p0}r_{NS}^{-1}$.
Therefore, the boundary layer luminosity is 
\beq
L_{BL}=\frac{\gamma}{\gamma-1}\dot M_{\rm Edd}c^2\frac{\mdot}{6r_{NS}}
\left(1-s^2\right)\approx 1.7\times10^{36}m\mdot_{-2}r_3^{-1},
\label{L-bl}
\eeq
where we have assumed $\gamma=5/3$. The total luminosity of the system
is $L=L_{SS}+L_{BL}$.

\subsection{Effect of Comptonization}
\label{S:PROP-COMPT}

Using the self-similar solution (\ref{sss}),(\ref{norm}), 
we may readily estimate the electron scattering optical depth and 
the $y$-parameter.\footnote{Here we just estimate where the effect
of Comptonization becomes significant. For better analytical 
approximations see, for instance, \citet{DLC91,TL95}. (In the 
latter paper, the expression for $y$ is not given, but it can be 
inferred using equation [24]:
$y=\tau_{es}[(\alpha+3)\theta/(1+\theta)+4d_0^{1/\alpha}\theta^2]$.)
Comptonization of free-free radiation has also been considered
by \citet{T88}.}
The optical depth is
\bea
\tau_{\rm es}&\simeq&\rho\kappa_{\rm es}R
\simeq10^3\alpha(1-\delta)^{1/2}(1-s^2)^{1/2}s^2r^{-1} \nonumber\\
&\sim&\alpha_{0.1}s_{0.1}^2r^{-1} ,
\eea
where $\kappa_{\rm es}=\sigma_T/m_p$ is the electron scattering opacity for
ionized hydrogen. Since $r\ge3$, we see that $\tau_{\rm es}\le1/3$ for
reasonable parameters and the radiation is optically 
thin to electron scattering. The $y$-parameter is
\bea
y&=&16\theta_e^2\tau_{\rm es}
\simeq2\times10^6\alpha(1-\delta)^{-1/2}(1-s^2)^{3/2}s^2r^{-2} \nonumber\\
&\sim&2\times10^3\alpha_{0.1}s_{0.1}^2r^{-2}.
\eea
The radius at which $y\sim1$ is
\beq
r_c\sim45\alpha_{0.1}^{1/2}s_{0.1} .
\eeq
Above this radius the inverse Compton scattering is small and the 
self-similar solution is valid. For $r<r_c$, however, Comptonization is 
important and the electron temperature profile will be modified from 
the self-similar form. Since the electron-proton collisions are relatively 
weak (the plasma is two-temperature), other quantities, e.g., the density, 
proton temperature, etc., are unaffected. Comptonization is 
unimportant for low-viscosity flows, $\alpha\la0.01$ around slowly 
rotating NSs, $s\la0.01$, because then $r_c<r_{NS}$.

\subsection{Spectrum}

We now estimate the spectrum of radiation emitted from the settling 
accretion flow. Let us neglect inverse Compton scattering for the moment.
The relativistic bremsstrahlung emissivity is 
approximated as $\epsilon_\nu\propto\rho^2\exp{-(h\nu/kT_e)}
\textrm{ erg cm}^{-3}\textrm{ s}^{-1}\textrm{ Hz}^{-1}$. 
Therefore the luminosity per unit frequency is
\bea
L_\nu&\propto&\int_{R_{NS}}^\infty\rho^2e^{-h\nu/kT_e}2\pi R^2\,dR \nonumber\\
&\propto&\int_{1/\nu_m}^\infty t^{-3}e^{-\nu t}dt
\propto\nu^2\Gamma(-2,\nu/\nu_m),
\eea
where $\Gamma(a,z)=\int_{z}^\infty t^{a-1}e^{-t}dt$ is the incomplete
gamma-function and $\nu_m=kT_e(R_{NS})/h$ is the maximum frequency.
Above $\nu_m$ the spectrum falls exponentially and below $\nu_m$
it is nearly flat. We may, thus, replace the exponential in the integral with
a square function which is equal to unity for $\nu<\nu_m$ and 0 for 
$\nu>\nu_m$. With this approximation
\beq
L_\nu\simeq\frac{3}{2}\frac{L_{SS}}{\nu_m}\left(1-\frac{\nu^2}{\nu_m^2}\right),
\eeq
where $L_{SS}=\int L_\nu d\nu$ is the total luminosity of the self-similar flow,
represented by equation (\ref{L-settl}). The break frequency,
$\nu_m$, is roughly given by  $h\nu_m\sim2.7\textrm{ MeV}$ for a typical 
electron temperature $T_{e,{\rm max}}\sim10^{10.5}~^\circ\textrm{K}$ 
[cf., equation (\ref{norm})]. 
At a typical x-ray energy, $h\nu\sim3\textrm{ keV}$, the observed
luminosity per decade is
\beq
\nu L_\nu\simeq1.7\times10^{31}m\alpha_{0.1}^2s_{0.1}^4
\left(\frac{h\nu}{3\textrm{ keV}}\right)\ \textrm{ erg s}^{-1},
\label{ss-spec}
\eeq
i.e., $\nu L_\nu\sim1.5\times10^{32}$ for a 300~Hz neutron star 
($s_{0.1}\sim1.6$). The luminosity per decade is much 
greater at higher photon energies and may be as high as 
$\sim\textrm{few}\times10^{34}-10^{35}\textrm{ erg/s}$ at 
$h\nu\sim\textrm{ MeV}$.

As shown in the previous section, Comptonization becomes important
below the radius $r_c$. At $r_c$, $y\approx1$ and the electron temperature is
\beq
T_e(r_c)\sim2.7\textrm{ MeV}/\sqrt{r_c}
\sim400\alpha_{0.1}^{-1/4}s_{0.1}^{-1/2}\textrm{ keV}.
\eeq
For $r<r_c$, the electron temperature will be determined 
self-consistently by Compton cooling rather than by bremsstrahlung
emission. Computing the spectrum from this region is beyond the
scope of the paper.  We also do not attempt to calculate the spectrum
of the radiation from the boundary layer.

\subsection{Bernoulli parameter}

It is known that the Bernoulli parameter of the accreting gas in BH
ADAFs is positive for a wide range of $r$ \citep{NY94,NY95a,NKH97}, 
and it has been suggested that the positive Bernoulli parameter may trigger 
strong winds or jets in these systems \citetext{\citealp{NY94,NY95a,BB99},
but see \citealp{ALI00}}.
\citet{IA99,IA00} confirmed with numerical simulations that strong 
outflows are produced from BH ADAFs when $\alpha\sim1$.

Normalizing the Bernoulli parameter, $Be$, by $(\Omega_KR)^2$,
and using equations (\ref{sss}),(\ref{norm}),
we find that the self-similar settling flow has
\bea
b\equiv {Be\over\Omega_K^2R^2}
&=&\frac{1}{v_k^2}\left(\frac{1}{2}v^2+\frac{1}{2}\Omega^2R^2-\Omega_K^2R^2
+\frac{\gamma}{\gamma-1}c_s^2\right) 
\nonumber\\
&=&\frac{v_0^2}{c^2}\,r+\frac{s^2}{2}-1+\frac{\gamma}{\gamma-1}\frac{1-s^2}{3}
\nonumber\\
&\simeq&-\frac{2\gamma-3}{3(\gamma-1)}
-\frac{s^2}{2}\frac{3-\gamma}{3(\gamma-1)},
\label{Bern}
\eea
where $\gamma$ is the mean adiabatic index of the flow and in the last 
expression we have neglected the term
in $v_0^2$ since it is negligible deep inside the settling region. 

We see that the second term in the final expression is always
negative.  This term is proportional to $s^2$, which means that a more
rapidly spinning NS stabilizes the accretion flow against outflows
more effectively than a slower spinning star.  The physical
explanation is as follows.  The centrifugal force increases with
increasing $s$, which has the effect of making it easier for the gas
to escape.  At the same time, however, the centrifugal force causes the radial
infall velocity to decrease, which increases the time available for
cooling. The temperature and the gas pressure go down, making the gas
more gravitationally bound. The net contribution of the two effects
turns out to be negative.

The first term in the last line of equation (\ref{Bern}) can be either
positive or negative, depending on the value of $\gamma$.  Combining
the two terms, we find that the gas is gravitationally bound and
unable to flow out in a wind (i.e. $b<0$) if the adiabatic index
satisfies 
\beq 
\gamma>\frac{3\left(1-\frac{1}{2}s^2\right)}{2-\frac{1}{2}s^2}.  
\label{b<0}
\eeq 
For $s^2\ll1$, the condition is $\gamma>1.5$; that is, the the accretion 
flow can produce a wind and/or a collimated outflow only if $\gamma<1.5$ and 
is stable to such outflows if $\gamma>1.5$.
Normally, we expect $\gamma$ to be close to 5/3 for the accreting gas.

\subsection{Stability to Convection}
\label{S:PORO-CONV}

It is well known that if the entropy increases inwards in a
gravitationally-bound non-rotating system, the gas is
convectively unstable; otherwise the flow is stable.  The specific
entropy profile in the settling accretion flow around a NS can be readily
calculated from equations (\ref{energy-p}),(\ref{energy-e}) using
(\ref{sss}),(\ref{norm}).  This gives 
\beq
\frac{ds}{dR}=\frac{k}{m_p}\frac{1}{\gamma-1}\frac{d}{dR}
\ln\left(\frac{c_s^2}{\rho^{\gamma-1}}\right)
=\frac{k}{m_p}\frac{2\gamma-3}{\gamma-1}\frac{1}{R}.  
\eeq 
We see that the entropy
increases outwards for $\gamma>1.5$ and inwards for $\gamma<1.5$.
Hence if $\gamma>1.5$ the flow is stable against convection, while if
$\gamma<1.5$ the flow is convectively unstable.

In the presence of rotation, the analysis is a little more complicated.
\citet{NIA99} and \citet{QG99} discuss the generalization of the 
Schwarzchild criterion for accretion flows with rotation. 
If the gas motions are restricted to the equatorial plane
of a height-integrated flow, convective 
stability requires the following effective frequency to be positive:
\beq
N^2_{\rm eff}=N^2+\kappa^2>0,
\eeq
where $N$ is the Brunt-V\"ais\"al\"a frequency and $\kappa$ is the epicyclic 
frequency, $\kappa=\Omega$ for $\Omega\propto R^{-3/2}$. For a power-law
flow with $\rho\propto R^{-a}$ and $\Omega(R)=s\Omega_K\propto R^{-3/2}$
with $s^2=1-(1+a)c_0^2$, this criterion may be written as follows
\citetext{see \citealp{NIA99} for more discussion}
\beq
N^2_{\rm eff}=\Omega^2_K\left(-[(\gamma+1)-a(\gamma-1)]
\frac{(1+a)c_0^2}{\gamma}+1\right)>0.
\label{Neff}
\eeq
Since for the self-similar settling solution $a=2$, the stability criterion 
(\ref{Neff}) becomes
\beq
N_{\rm eff}^2=\frac{\Omega_K^2}{\gamma}
\left[(2\gamma-3)+s^2(3-\gamma)\right]>0
\eeq
which yields that the flow is convectively stable if
\beq
\gamma>\frac{3(1-s^2)}{2-s^2}.
\label{c>0}
\eeq 
This condition is different from the stability criterion against outflows,
given in equation (\ref{b<0}).

Following the techniques developed by \citet{QG99}, \citet{NIA99} have
also presented a more general analysis of convection in a self-similar
accretion flow.  This analysis, which does not restrict motions to lie
in the equatorial plane, assumes that $v_\phi$ and $c_s$ are
independent of the polar angle $\theta$ \citetext{as is valid for a marginally
convectively stable system, cf \citealp{QG99}}.  \citet{NIA99} find
that the most unstable region of the flow is near the rotation axis,
$\theta=0,\pi$.  They show that the marginal stability criterion for
this polar fluid coincides with the condition for the positivity of
the Bernoulli parameter.  That is, a flow which is convectively stable
at all $\theta$ has a negative Bernoulli parameter, while a flow which
is convectively unstable for at least some values of $\theta$ has a
positive Bernoulli parameter.  (The Bernoulli parameter itself is
independent of $\theta$.)

We have verified this result for the solutions presented in this
paper.  Specifically, when we apply to our solution the more general convective
stability criterion given by equation (A9) of \citet{NIA99}, 
we recover the condition (\ref{b<0}) above, namely that the
self-similar flow is convectively stable if and only if 
\beq
\gamma>\frac{3\left(1-\frac{1}{2}s^2\right)}{2-\frac{1}{2}s^2}.
\label{convstable}
\eeq

\section{Comparison with Numerical Results}
\label{S:NUM}

We have numerically solved the system of height-integrated
two-temperature fluid equations (\ref{mdot})--(\ref{energy-e}) with
boundary conditions.  In the energy equations we assume that viscous
dissipation only heats the protons (i.e. $\delta=0$). We include
energy transfer from protons to electrons via Coulomb collisions, and
we take the cooling of electrons to be purely by free-free emission,
as discussed in \S \ref{S:SSS}. For these processes
we use the expressions given in
\citet{NY95}, which smoothly interpolate between the regimes of
non-relativistic and relativistic electrons.  We take into account the
variation of the electron adiabatic index $\gamma_e$ with temperature
\citep{Chandra39,EMN97}, using a simple interpolation formula from
\citet{GP98}.  We assume that the protons have $\gamma_p=5/3$.

We employ the gravitational potential of \cite{PW80} to mimic the
effect of strong gravity near the NS surface.  In this potential the
Keplerian angular velocity takes the form 
\beq
\Omega_K^2=\frac{GM}{(R-R_S)^2R}.  
\eeq 
Note that the analytical work presented in the previous sections 
is based on a Newtonian potential.

We specify the boundary conditions as follows.  We take the outer
boundary of the flow to be at $r_{\rm out}=10^6$.  At this radius we
specify that the angular velocity is equal to its value in the
self-similar ADAF solution of \citet{NY94}, and that the proton and
electron temperatures are both equal to the self-similar ADAF
temperature.  We assume that the accreting star is a $1M_{\sun}$
neutron star with a radius $R_{NS}=3R_S=8.85$ km, unless stated otherwise.  
At $R=R_{NS}$, we specify the value of the NS spin parameter,
$s=\Omega_{NS}/\Omega_K(R_{NS})$, and we require the proton
temperature of the flow to be $T=\mathrm{few}\times10^7\textrm{ K} \ll
T_{virial}$.  (We do not assume that the electron and proton
temperatures are equal, but in fact they are equal.)  We do not
constrain the density of the gas in any way at either boundary.

The numerical problem as posed here has a family of solutions
characterized by three dimensionless parameters: the mass accretion
rate $\mdot$ (in Eddington units), the NS spin $s$ (in units of the
Keplerian angular velocity at the NS surface), and the viscosity
parameter $\alpha$.  The angular momentum flux $\dot J$, or
equivalently the dimensionless flux $j=\dot J/\dot
M\Omega_K(R_{NS})R_{NS}^2$, is an eigenvalue of the problem.

Figure \ref{f:profile} shows representative solutions for $\alpha=0.1$
and a range of values of $\dot m$ and $s$.  The solutions clearly have
three radial zones.  For $r>10^{2.5}$, there is a one-temperature zone
in which the gas properties vary roughly as power-laws of the radius.  For
$r<10^{2.5}$, there is a second power-law zone with a two-temperature
structure.  Finally, close to the NS, the flow has a boundary layer
region.  In this final region, the gas experiences run-away cooling,
the velocity falls precipitously, and the density increases very
rapidly.  This region of the flow does not have power-law behavior.

The numerical solutions are unreliable in the boundary layer; thermal
conduction (not included in the calculations) is probably very
important here, and optical depth effects (also not included) will
modify the radiation properties significantly.  The solutions are
suspect also in the inner region of the two-temperature power-law
zone, where Comptonization is likely to be important.  Outside these
regions, however, the numerical solution is expected to be accurate.

The numerical results in the two-temperature power-law zone below
$r\sim10^{2.5}$ agree quite well with the analytical solution
presented in \S\ref{S:SSS-I}.  Curves corresponding to a given value
of $s$ and different values of $\mdot$ coincide with one other to very
good accuracy, as predicted by the analytical solution.  This is best
seen in the profiles of $\rho$ and $\Omega$.  Changing $s$ causes an
up/down shift of the curves but does not affect the slopes of the
curves. The temperature profiles are sensitive to the spin $s$,
especially for large values of $s$.  The radial velocity varies
approximately as $v\propto\mdot$ and is roughly consistent with
$v\propto r^0$ for $s>0.1$.

For $r>10^{2.5}$, the numerical solutions are in reasonable agreement
with the one-temperature self-similar solution described in \S
\ref{S:SSS-O}.  The agreement is less perfect than in the previous
zone.  This is primarily because the analytical solution requires a
very low value of $\dot m$ in order to be valid as far out as the
outer radius $r\sim10^6$ [cf., equation (\ref{out-constr})].  
The numerical models shown have
larger values of $\dot m$ than this limit.

As we discussed in \S\ref{S:SSS} and \S\ref{S:PROP}, the transport of
angular momentum in a hot settling flow differs dramatically from the
well-known behavior of a thin disk.  The solid line in Figure
\ref{f:j-vs-s} indicates the dependence of the dimensionless angular
momentum eigenvalue $j$ as a function of the dimensionless NS spin
$s$.  The long-dashed line indicates the corresponding results for a
thin disk \citep{PN91}, where $j\simeq+1$ for most values of
$s$, and goes negative only for stars nearly at break-up.  In
contrast, in the settling flow, $j$ is negative for almost all values
of $s$.  For the particular choice of parameters, namely $\alpha=0.1,\
\mdot=0.01$, $s\sim\textrm{few}\times10^{-1}$, we find that $j\sim-$
few.

The short-dashed line in Figure \ref{f:j-vs-s} shows the analytical
formula for $j$, as given in equation (\ref{j}).  The agreement with the
numerical results is good for a wide range of $s$ below about 0.5.
For $s>0.5$, the numerically determined $j$ levels off at a constant
negative value, whereas the analytical result shows $|j|$ decreasing
rapidly.  The main reason for the discrepancy is the neglect of the
ram pressure term in the radial momentum equation in the analytical
work.  Note the interesting fact that super-Keplerian accretion
($s>1$) is, in principle, possible (provided one can arrange to have a
star with super-Keplerian rotation).  In a super-Keplerian flow, the
ram pressure of the infalling gas supplies the radial momentum needed
to push the gas onto the NS.  For extremely small $s\ll0.1$, the
numerical solutions show $j$ to be slightly positive; we find
$j\sim10^{-3}$, as indicated in the lower panel in Figure
\ref{f:j-vs-s}.  Here again the self-similar solution, which predicts
a small negative value for $j$, breaks down.  For very small $s$, the
$\dot J$ term in equation (\ref{omega}) is comparable to or smaller
than the $\dot M\Omega R^2$ term which was omitted in deriving the
analytical solution.  This is the reason for the discrepancy.  The
precise value of $s$ at which the analytical solution breaks down
depends on the choice of parameters ($\alpha$, $\mdot$), but is 
essentially independent of the outer radius. Thus, the transport of 
angular momentum through the flow and the spin-down of the star are 
determined by the boundary conditions at the stellar surface, 
and not by the outer boundary of the flow.

One of the most surprising features of the self-similar solution is
that the angular momentum flux $\dot J$ is independent of $\dot m$;
equivalently, the dimensionless eigenvalue $j$ is $\propto \dot
m^{-1}$.  The solid line in Figure \ref{f:j-vs-mdot} is a plot of $-j$
as a function of $\mdot$ for a flow with $\alpha=0.1$ and $s=0.3$, as
determined from the numerical solutions.  The dashed line indicates
for comparison the analytical scaling, $j\propto \dot m^{-1}$, with
the proportionality constant given in equation (\ref{j}).  Note the
very good agreement between the numerical results and the analytical
self-similar solution.

The highest value of $\dot m$ up to which we could obtain a numerical
solution is $\mdot_{\rm crit}=0.0313$.  Beyond this critical value,
there is no hot solution.  (The value of $\mdot_{\rm crit}$ depends on
$s$, $\alpha$ and $r_{out}$.)  For larger $\mdot$, the density in the
flow is so high that there is runaway free-free cooling and the gas is
unable to remain hot.  We presume that the accretion then occurs via a
thin accretion disk.

Figure \ref{f:lumin} plots, for selected values of $\mdot$, the
luminosity per logarithmic interval of $D$, where $D$ is the
fractional distance from the NS surface:
\beq 
D={(R-R_{NS})\over R_{NS}}.  
\eeq 
We see that the luminosity at the peak of the curve is
very insensitive to $\mdot$.  The emission in the peak corresponds to
radiation from the settling flow.  This emission represents energy
released by the spin-down of the NS, and its luminosity is independent
of $\mdot$ [cf. equation (\ref{L-settl})].  For radii below the peak, the
curves do show a dependence on $\mdot$.  The radiation here
corresponds to boundary layer emission, which is proportional to
$\mdot$ according to equation (\ref{L-bl}).  

All the models described above have $R_{NS}=3R_S$. However, 
different equations of state predict slightly different NS radii,
$R_{NS}=2-4R_S$. We have computed numerical models for this range 
of $R_{NS}$ and we find that the subsonic settling solution exists 
for the whole range. Qualitatively, the solutions with different $R_{NS}$
are very similar. As $R_{NS}$ decreases, the peak temperature is higher,
as expected for the deeper potential.

\section{Relationship of the Settling Flow to an ADAF}
\label{S:TRANS}

The accretion solution we have discussed so far radiates all the
energy dissipated by viscosity, and is therefore
``cooling-dominated.''  On the other hand, it is known that a hot
flow around a black hole is an ``advection-dominated''
accretion flow (ADAF).  Both solutions have a two-temperature
structure for $r\la10^{2.5}$ and both are very hot (nearly virial) for
all $r$.  How are these two types of accretion flows related to each
other?

By solving the equations numerically for different boundary
conditions, we have found that the two solutions are part of a single
sequence of solutions in which the spin of the star, $s$, plays a
pivotal role as a control parameter.  For relatively rapidly rotating
stars, with $s\ga0.1$, we obtain the settling solution in our
numerical experiments.  However, as $s$ is decreased, we find that the
settling solution smoothly transforms to an ADAF-type solution, which
becomes well-established for $s\la0.01$.  The transition is not sharp, so
it is difficult to identify a specific transition point $s=s_t$ at
which the transformation occurs.  Numerical experiments indicate that
the value of $s_t$ (however it is defined) is not very sensitive to
$R_{out}, \gamma$, and $\mdot$ and is, roughly, $s_t\sim0.04-0.06$.

The change of the nature of the flow as $s$ is varied is illustrated
in Fig.\ \ref{f:profiles2}.  The solid and dotted curves correspond to
two solutions with $s=0.3$ and $s=0.01$, respectively, with all other
boundary conditions being the same.  We see that the solutions are
markedly different from each other.  This is most clearly seen in the
profiles of density, where the $s=0.3$ model has a logarithmic slope
of -2, as appropriate for the cooling-dominated settling solution
described in this paper, and the $s=0.01$ model has a slope of -3/2,
as expected for a standard self-similar ADAF \citep{NY94,NY95a}.
There is a similar difference also in the profiles of the radial
velocity, where the two solutions have logarithmic slopes of -1/2 and
0, respectively.

An interesting feature of the $s=0.01$ ADAF-type solution is that it
consists of two distinct segments.  For large radii (in Fig. 
\ref{f:profiles2}, for radii outside $r\sim20$), the flow corresponds 
to the standard ADAF discussed in the literature, with the scalings 
\beq 
\rho\propto r^{-3/2}, \qquad c_s^2\propto r^{-1}, 
\qquad \Omega\propto r^{-3/2}, \qquad v\propto r^{-1/2}.  
\eeq 
However, at smaller radii, the
numerical solution indicates the presence of a second
advection-dominated zone, a ``settling ADAF,'' which was first seen in
numerical calculations described in \citet{NY94}.  This settling ADAF
is seen in Fig.\ \ref{f:profiles2} as a zone that lies between the
boundary layer region and the outer standard ADAF, with different
slopes for $\rho$ and $v$.  The radial extent of the settling ADAF
zone may be quite large and, in general, depends on the flow
parameters and boundary conditions.

A self-similar model of the settling ADAF may be readily obtained as
follows. In an ADAF, energy is not radiated, therefore $q^-=0$. Close
to the star $\Omega\simeq\textrm{ constant}$, therefore $q^+=0$.
Equations (\ref{energy-p}), (\ref{energy-e}) then simplify to the
condition of entropy conservation, $ds/dR=0$, which yields
$c_s^2\propto\rho^{\gamma-1}$.  As the material settles on the star,
its radial velocity decreases and we have $v\ll v_{ff},\
\Omega\ll\Omega_K$.  Then, from equation (\ref{mom}), it follows that
the temperature of the gas is nearly virial. Other quantities are
determined straightforwardly, so that we have 
\beq 
\rho\propto r^{\frac{1}{\gamma-1}}, \qquad c_s^2\propto r^{-1}, \qquad
\Omega\sim\textrm{const.}, \qquad v\propto r^{-\frac{2\gamma-3}{\gamma-1}}.  
\eeq 
The infall velocity decreases
with radius if $\gamma<1.5$, and increases if $\gamma>1.5$.  To
highlight the difference between the standard ADAF and the settling
ADAF, we have chosen $\gamma=4/3$ in the solutions shown in Fig.\
\ref{f:profiles2}.

Finally, the long-dashed curves in Fig.\ \ref{f:profiles2} correspond
to a solution with $s=0.3$ for which we have increased the outer
boundary value of $T$ by a factor of 10.  We see that the solution in
the interior is not sensitive to the outer boundary conditions (within
a reasonable range, of course).

\section{Summary and Discussion}
\label{S:SandD}

In this paper, we have presented analytical and numerical solutions
that describe a hot, viscous, two-temperature accretion flow onto a
neutron star (NS).  The results are relevant also for accretion onto
other compact stars with a surface, e.g. white dwarfs.  To our
knowledge, this is the first study of viscous fluid dynamics for a hot
flow around a NS.

The presence of a surface modifies the nature of the flow relative to
the case of a black hole.  We show that the accretion flow has an
extended settling region in which the radial velocity $v$ is constant;
$v$ is also small relative to the local free-fall velocity.  The
density in the settling region varies as $\rho\propto r^{-2}$, and the
angular velocity has a Keplerian scaling, $\Omega=s\Omega_K \propto
r^{-3/2}$, with $s$ being a constant. Here, $r$ is the radius in
Schwarzchild units, and the value of $s$ is set by the spin of the
NS: $s=\Omega_{NS}/\Omega_K(r_{NS})$.  At the inner edge of the
settling region, there is a narrow boundary layer in which the
velocity falls extremely rapidly and the density increases sharply to
match the surface density of the NS.

The settling region consists of two distinct zones.  In the inner
zone, $r\lesssim 10^{2.5}$, the gas is two-temperature, with the proton
temperature varying as $T_p\propto r^{-1}$ and the electron
temperature varying as $T_e\propto r^{-1/2}$.  These scalings are
derived assuming that electrons radiate primarily by free-free
emission and that energy transfer from protons to electrons occurs via
Coulomb collisions.  We have derived a completely general analytical
self-similar solution for this region which agrees well with numerical results.

In the outer zone of the settling region, $r\gtrsim10^{2.5}$, the gas is
one-temperature, $T_p\approx T_e\propto r^{-1}$; here again, we derive
an analytical self-similar solution which agrees reasonably well with
numerical results.

The most surprising feature of the settling region is that nearly all
the gas properties are independent of the mass accretion rate $\mdot$;
only the radial velocity shows a dependence: $v\propto\mdot$.  Since
the density and temperature are independent of $\mdot$, the luminosity
is also independent of $\mdot$.  Indeed, the settling solution is
valid---with a finite luminosity---even in the limit when $\mdot\to0$.
Clearly, the luminosity does not originate from the gravitational
release of energy as mass accretes onto the NS.

The magnitude of the luminosity is very sensitive to the spin
parameter $s$ of the NS, varying as the fourth power of this quantity,
equation (\ref{L-settl}).  For $s\sim0.1$, as appropriate for the 
millisecond X-ray pulsar, SAX J1808.4-3658, the model predicts an X-ray 
(few keV) luminosity $\nu L_\nu\sim10^{32} ~{\rm erg\,s^{-1}}$ 
(\S\ref{S:PROP-LUMIN}).  This
estimate does not include the contribution from Comptonization in the
inner regions of the flow, which might increase the X-ray luminosity
by an undetermined amount.  We note that quiescent X-ray luminosities
of soft X-ray transients, including SAX J1808.4-3658, are generally in
the range $\nu L_\nu\lesssim10^{33} ~{\rm erg\,s^{-1}}$ 
\citep{NGMc97,Asai+99,Menou+99,Stella+00}.

The angular momentum flux $\dot J$ in the settling solution is
dominated by the viscous transport term rather than the advection term
$\dot M\Omega_KR^2$.  Consequently, $\dot J$ is negative, i.e. the
angular momentum flux is oriented outward, and the accretion flow
spins down the star.  The analytical solution predicts that spin-down
occurs for all values of the spin parameter $s$ of the central star.
The numerical solutions by and large confirm this; for plausible
parameters, spin-up is seen only for extremely small values of the
spin parameter, $s<0.005$ (cf Fig. \ref{f:j-vs-s}). (The exact value
of $s$ at which $\dot J$ changes sign depends on $\alpha$ and $\mdot$,
but is relatively independent of the position of the outer edge of the
flow.)  This behavior is very different from the case of a thin
accretion disk \citep{PN91,Paczynski91}, where one finds that the star
is spun-up for nearly all values of $s$, and spin-down occurs only for
$s$ close to unity (break-up limit).

Another surprising feature of the settling solution is that $\dot J$,
like nearly all other quantities, is independent of $\mdot$.  Indeed,
the settling zone behaves like a stationary zone (since $v$ is very
small), and essentially acts like a conventional ``brake,'' slowing
down the star by viscosity.  The brake can operates even if
$\mdot\to0$, so long as there is a static atmosphere of the
self-similar form and there is a sink for the angular momentum, say an
external medium, at large $r$.  Furthermore, the luminosity of the
settling flow is almost entirely from the energy released by the
viscous braking action.  That is, the luminosity ultimately is fed by
the loss of rotational kinetic energy of the star, and not by gravity.

This result has an interesting consequence.  In accretion flows around
black holes, one defines an efficiency factor $\eta$ by comparing the
accretion luminosity $L_{acc}$ to the rest mass energy of the
accreting gas, $\eta\equiv L_{acc}/\dot Mc^2$.  It is well-known that
$\eta=0.06$ for a thin accretion disk around a Schwarzchild black hole
and that the value increases to $\eta=0.42$ for a maximally rotating
Kerr hole.  A number of interesting ideas have been discussed in the
literature for increasing the efficiency of an accretion flow around a
black hole; these involve tapping the rotational energy of the black
hole using magnetic fields or viscosity
\citep{BlZnaek,Krolik99,Gammie99}. The general relativistic dragging
of inertial frames by the spinning hole plays an important role in the
mechanism of \citep{BlZnaek}.

For the hot settling solution described in this paper, the luminosity
is almost entirely from the rotational energy of the star.  Since
$L_{acc}$ is independent of $\mdot$, the efficiency scales as
$\eta\propto\mdot^{-1}$ and $\eta\to\infty$ as $\mdot\to0$.  Thus, it
would appear that accretion flows can tap the rotation energy of a
star with a surface more easily than the energy of a spinning black hole.  
\citet{SS00} showed that the extraction of rotational energy of a NS may 
result in very high boundary layer efficiencies, up to $\eta\sim0.67$, 
in the counter-rotating NS--thin disk systems, as well.
As in the case of the black hole, the energy extraction works best when 
the star is spinning rapidly: $\eta\propto s^4$.  Interestingly, the 
energy extraction is not a general relativistic effect --- our analytical
solution is based entirely on Newtonian physics.

\citet{YWV97} and \citet{YW98} recently suggested that the sudden
torque-reversal events seen in some accretion-powered pulsars may be
due to the accretion flow switching between a Keplerian thin disk and
a hot a sub-Keplerian state akin to an ADAF.  Our work lends support
to this suggestion.  We find that the torque does reverse in sign
between a thin accretion disk and a hot settling flow for almost any
reasonable stellar spin parameter.  We also find that the magnitude of
the spin-down torque exerted by the settling flow is comparable to the
measured value of the torque in torque-reversing pulsars
(\S\ref{S:PROP-SPIN}).

It is worth emphasizing that the settling solution described here is
quite distinct from the self-similar ADAF solutions derived for black
hole accretion \citep{NY94,NY95a,H96,KN98,BB99,Manmoto+00}. All the 
black hole solutions described in the literature have density varying
relatively mildly with radius: $\rho\propto r^{-3/2}-r^{-1/2}$.  Our
settling solution has $\rho\propto r^{-2}$.  Also, the black hole
solutions are advection-dominated, whereas our solution radiates
essentially all the energy it generates through viscous dissipation.
There are also, as we now discuss, significant differences in the sign
of the Bernoulli parameter and in the stability to convection.

\citet{NY94,NY95a} showed that their self-similar ADAF solution has a
positive Bernoulli parameter so long as the adiabatic index $\gamma$
of the gas is less than $5/3$.  They argued on the basis of this that
ADAFs are likely to have strong outflows and winds \citetext{but see
\citealp{ALI00}}.  Such strong outflows were confirmed with numerical
simulations by \citet{IA99,IA00}; they found outflows for large values
of the viscosity parameter: $\alpha\sim1$. \citet{BB99} developed a
self-similar model with inflow and outflow (the ADIOS model).  For the
settling solution described in this paper, the Bernoulli parameter is
positive only if $\gamma$ is less than $\gamma_{crit}$, where
$\gamma_{crit}=1.5$ for a slowly-spinning star and is smaller than 1.5
for a rapidly-spinning star [cf equation (\ref{b<0})].  Since the hot
ionized two-temperature gas in the flow is likely to have $\gamma$
close to $5/3$ at most radii, we expect the Bernoulli parameter to be
generally negative.  Therefore, we do not expect a strong outflow.  Of
course, this conclusion assumes that we do not have dynamically
important magnetic fields in the flow.

\citet{NY94,NY95a} also showed that their ADAFs are
convectively unstable for a wide range of conditions.  The convective
instability has been seen in numerical simulations in which the
viscosity parameter is assigned a low value: $\alpha\lesssim0.1$
\citep{ICA96,IA99,IA00,SPB99,IAN00}. Self-similar solutions for 
convection-dominated accretion flows (CDAFs) have been derived by 
\citet{NIA99} and \citet{QG99}. For the settling
solution described in this paper, we find that the gas is convectively
unstable only for the same low values of $\gamma$ for which the
Bernoulli parameter is positive.  (\citealp{NIA99} showed that, in general, 
for self-similar flows the criterion for the Bernoulli parameter to be
positive is the same as the criterion for the flow to be convectively
unstable, cf. \S\ref{S:PORO-CONV}.)  Thus, we do not expect hot settling 
flows around NSs to be convectively unstable, or to have a distinct CDAF 
mode of settling.

From numerical experiments we have discovered the interesting property
that the settling flow can continue well inside the last stable orbit
(down to at least $r_{NS}\sim2$) and yet remain subsonic at all radii.
This is despite the fact that the numerical models employ a
pseudo-Newtonian potential which mimic the last stable orbit for test
particles.  We find that the structure of the flow is qualitatively
similar for flows with $r_{NS}<3$ and $r_{NS}>3$.  This is very
different from black hole ADAFs which become supersonic close to the
central object.

We have also shown (\S5) that the settling solution and the ADAF are
not two physically distinct solutions, but are related to each other.
As the neutron star spin is decreased, we find that the settling flow
smoothly transforms to an ADAF-type solution.  The transformation
proceeds over the spin range $s\sim0.01-0.1$.

The settling solution is hot but cooling-dominated.  It is thus most
closely related to the two-temperature solution discovered by Shapiro,
Lightman \& Eardley (1976).  The SLE solution is known to be thermally
unstable (Piran 1978; Wandel \& Liang 199?; Narayan \& Yi 1995b), so
one may wonder about the thermal stability of our settling solution.
We defer discussion of this important topic to a future paper.

Could there be non-settling solutions around NSs, and could such flows
be more analogous to the black hole ADAF, ADIOS, and CDAF solutions?  Indeed
this is possible if the NS radius is small enough.  We could imagine,
for instance, a standard black hole-like flow around a NS, which makes
a sonic transition to a supersonic state, and then stops suddenly at a
standing shock at the surface of the NS.  Such a solution would be
dynamically consistent, and, except for the shock, would be very
similar to a black hole flow.  However, for such a solution to exist,
one requires the radius of the NS to be smaller than the sonic radius
of the flow.  The latter radius is estimated to be in the range
$r_{sonic} \sim2-5$, depending on the value of $\alpha$
\citep{NKH97,PG98}, whereas NS radii
are in the range $r_{NS}\sim 2-4$ for typical NS equations of state.
Thus,  for some choices of $\alpha$ and some
equations of state, $r_{NS}$ could be smaller than $r_{sonic}$.  In
such cases, we could have four different hot solutions around a NS: (i) we
could have the self-similar settling solution of the kind presented in
this paper (as we discussed in \S\ref{S:NUM}, we find subsonic settling
solutions for any choice of $R_{NS}$ in the range $2-4R_S$);
or (ii) we could have a \citet{NKH97}--like ADAF
solution with a shock at the NS surface; or (iii) we could have a
\citet{BB99}--like ADIOS solution with a shock at the
NS surface; or (iv) we could have a \citet{NIA99} and \citet{QG99}--like 
CDAF, again with a shock at the NS surface.
In the latter three cases, we expect the NS to be {\it spun-up} rather
than spun-down by the accretion flow, and we also expect the radiative
efficiency to be close to the standard value for a NS, namely
$\eta\sim0.1-0.5$.  The spectrum of the radiation is
also likely to be very different in the four models.  This may provide
a way to distinguish which if any of these possibilities is found in
nature.

In addition to the above possibilities, yet other flow configurations
may be possible when we allow for the multi-dimensional nature of the
flow.  These could be explored with numerical hydrodynamics
simulations.

\acknowledgements 

This work was supported in part by grants PHY~9507695 and AST~9820686 from
the National Science Foundation.


\figcaption[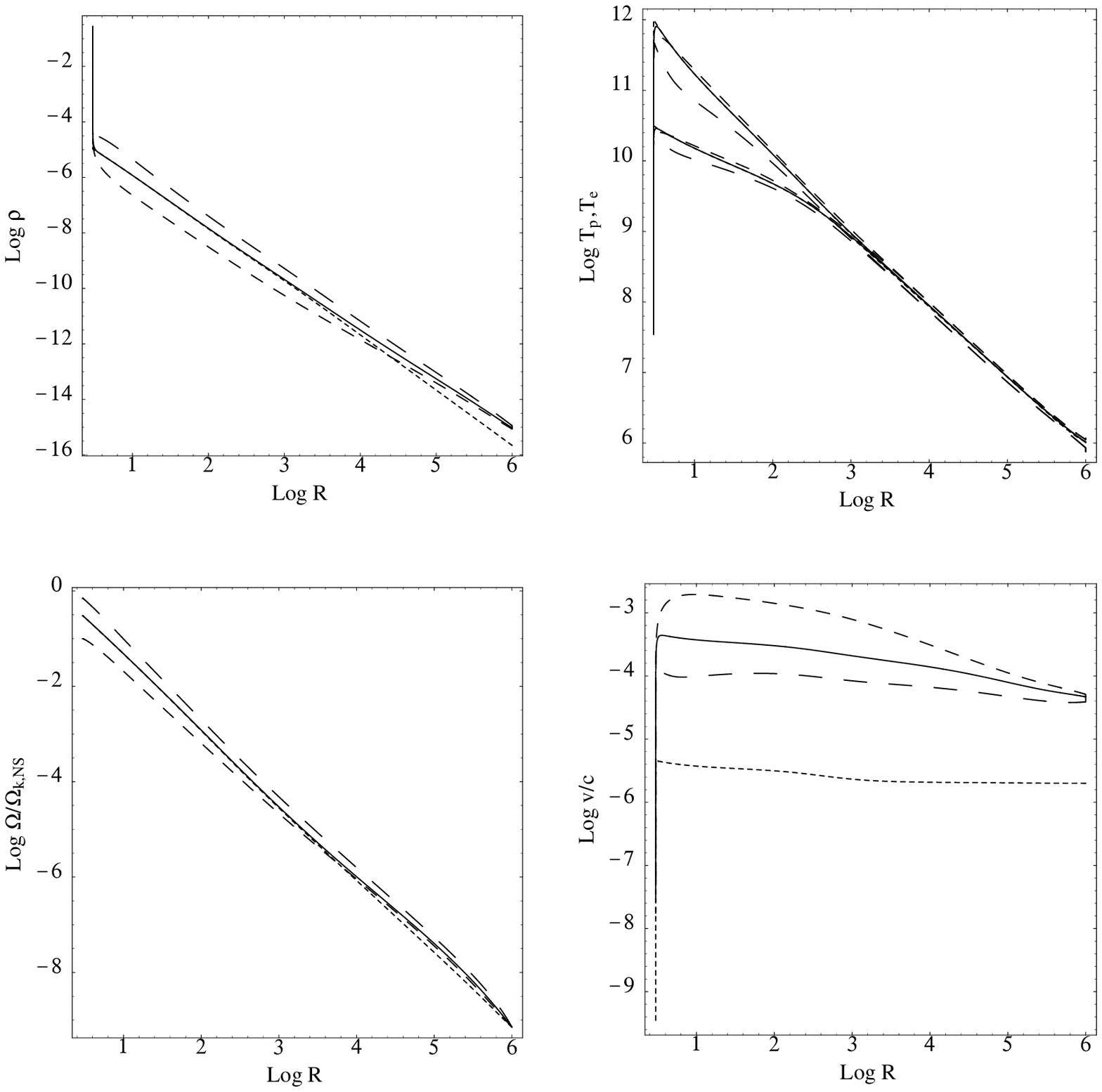]{Profiles of density $\rho$ (g~cm$^{-3}$),
proton temperature $T_p$ ($^\circ$K), electron temperature $T_e$
($^\circ$K), angular velocity $\Omega$ (in units of the Keplerian
angular velocity at the NS radius $R_{NS}$), and radial velocity $v$ (in
units of $c$) for accretion flows with $\alpha=0.1$ and $(\mdot,s$) =
(0.01,0.3) -- solid line, (0.0001, 0.3) -- short-dashed line,
(0.01,0.1) -- medium-dashed line, (0.01,0.7) -- long-dashed line.
\label{f:profile} }

\figcaption[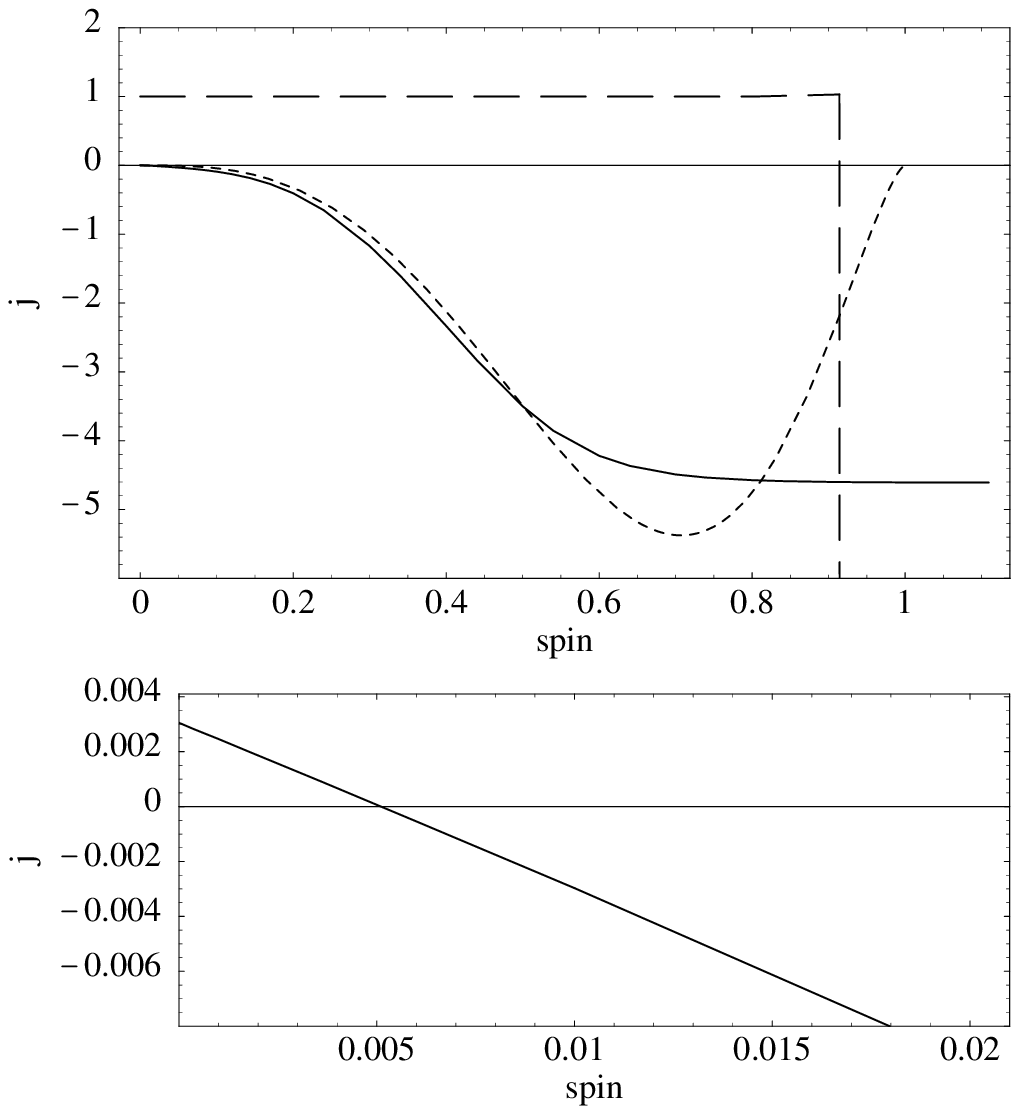]{The solid curve corresponds to the
dimensionless specific angular momentum flux $j$, shown as a function
of the dimensionless spin of the NS $s$, for $\mdot=0.01$ and
$\alpha=0.1$. The long-dashed line is $j(s)$ for the thin disk case,
taken from \citet{PN91}. The short-dashed curve corresponds to the
analytical self-similar solution, equation (\ref{j}).  The lower panel
gives a close-up of the small-$s$ region of the upper plot.
\label{f:j-vs-s} }

\figcaption[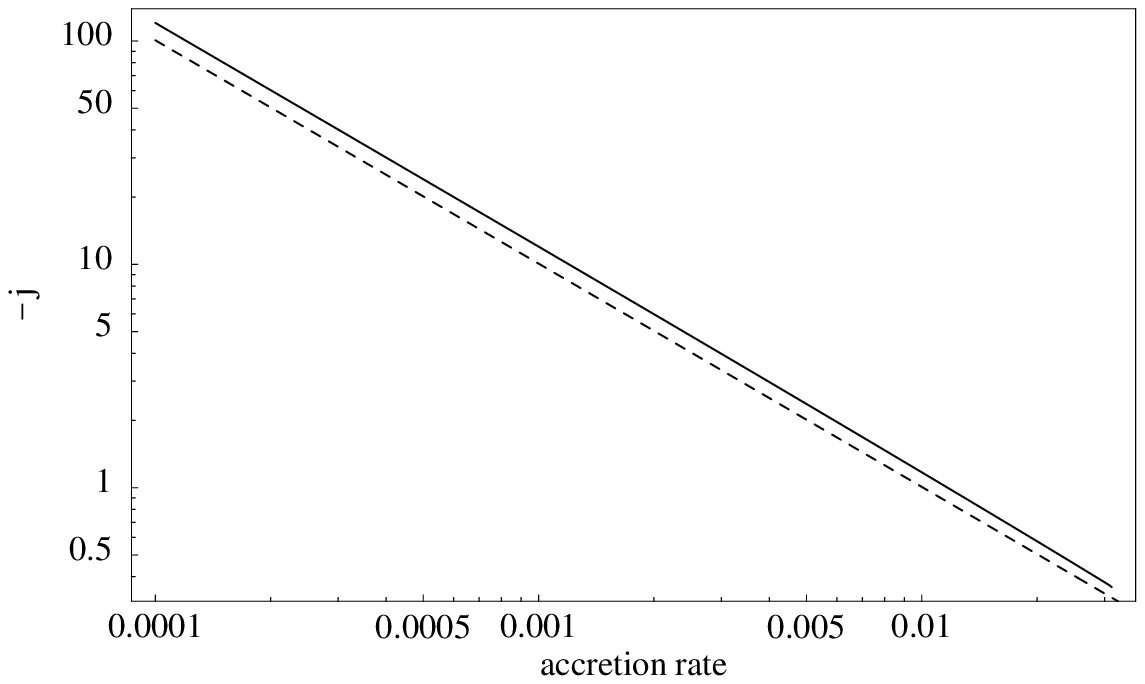]{Variation of the angular momentum flux
parameter $-j$ as a function of $\mdot$ for $s=0.3$, $\alpha=0.1$.
The solid curve corresponds to the results from numerical
computations, and the dashed curve corresponds to the analytical
self-similar solution.
\label{f:j-vs-mdot} }

\figcaption[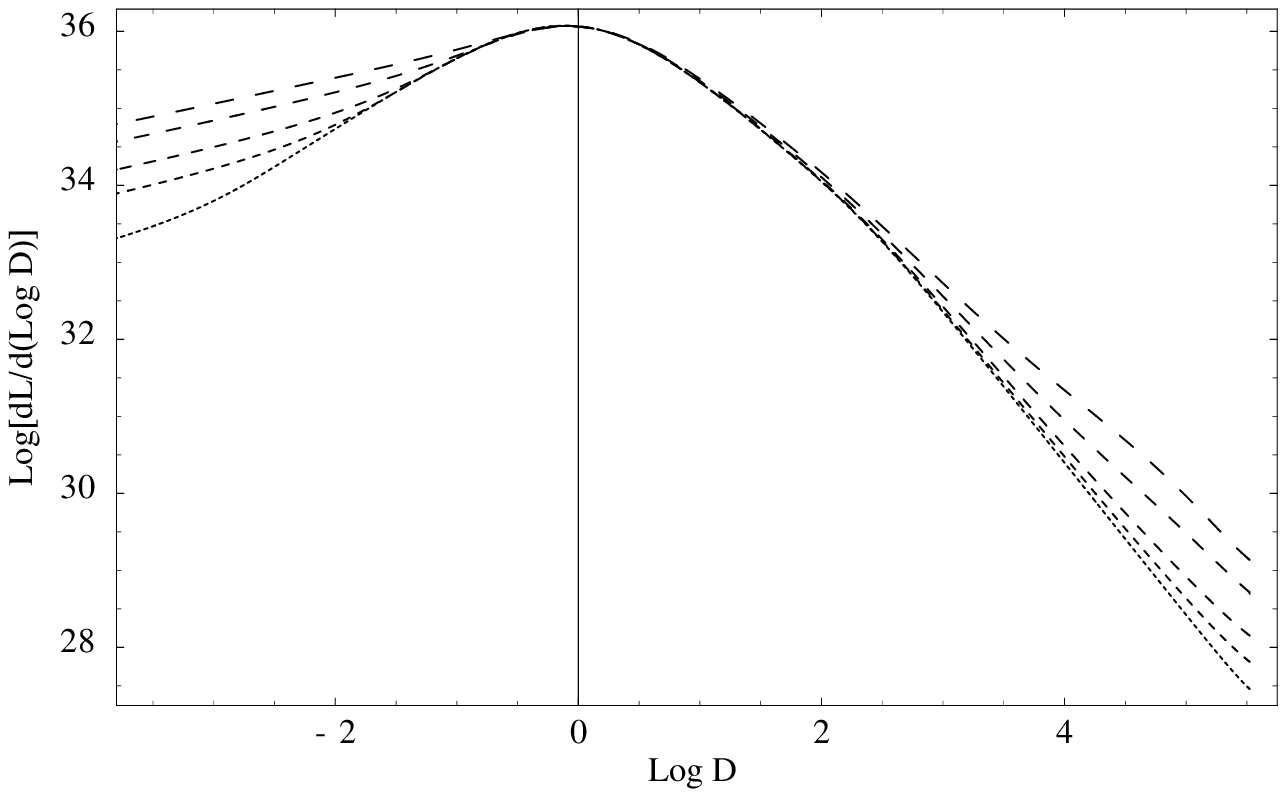]{Luminosity (erg/s) of the settling flow per
unit logarithmic interval of $D=(R-R_{NS})/R_{NS}$ for $\alpha=0.1,\
s=0.3$.  The curves correspond to different choices of the mass
accretion rate: from below, $\mdot= 0.0001,\ 0.001,\ 0.003,\ 0.01,\
0.02$.  A substantial part of the luminosity is independent of
$\mdot$; this part is from the settling flow.  There is some weak
dependence of the luminosity on $\mdot$ at small $D$, due to boundary
layer emission, and at large $D$, due to non-self-similar behavior of
the numerical solution.
\label{f:lumin} }

\figcaption[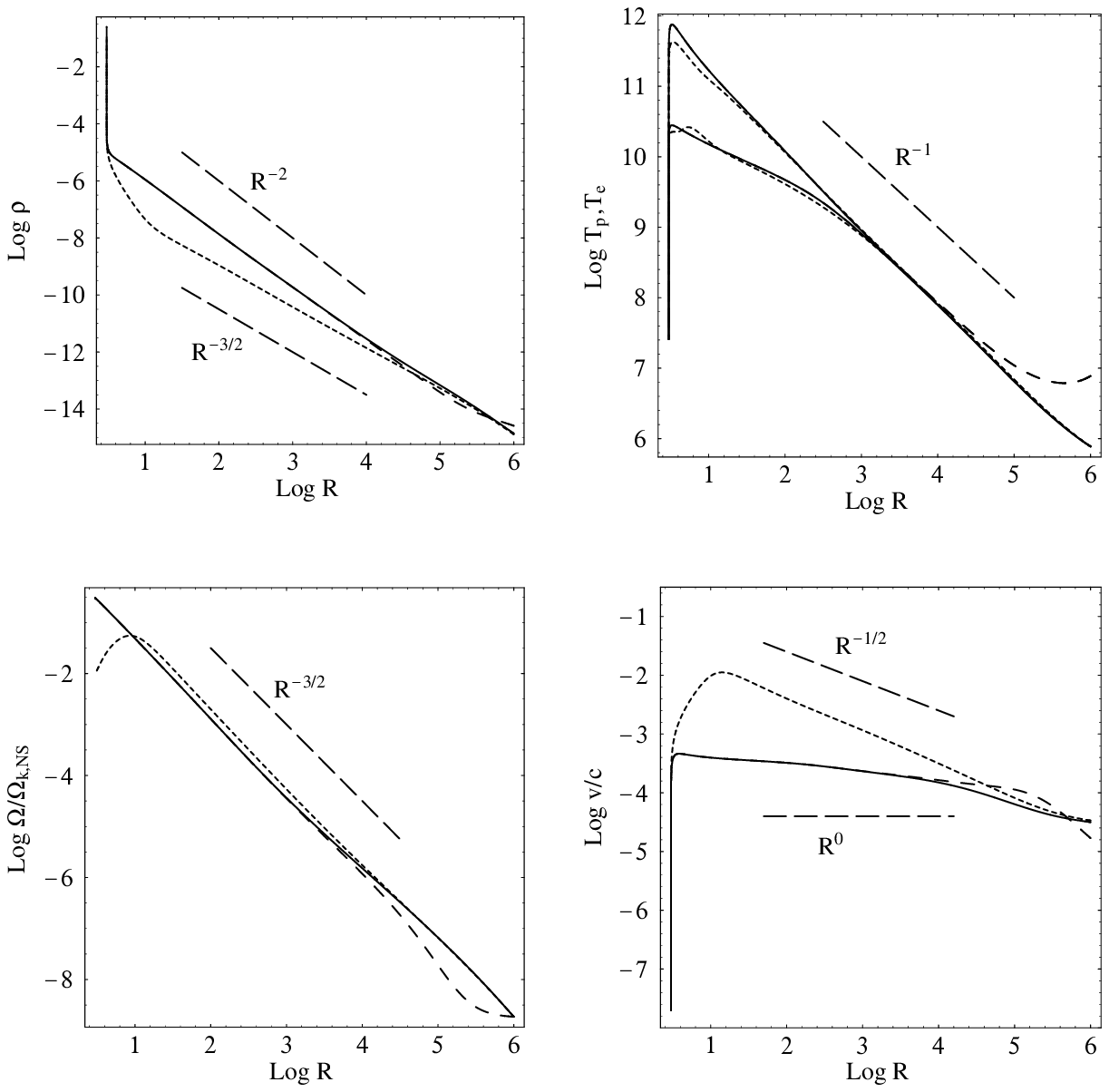]{Same as in Fig.\ \ref{f:profile} for
$\gamma=4/3$ and $s=0.3$ (solid curve) and $s=0.01$ (dotted curve).
Corresponding self-similar slopes for an ADAF and the settling flow
are also show.  The long-dashed curves represent the same solution as
the solid curve, but with ten times higher temperature at $R_{out}$.
\label{f:profiles2} }

\newpage
\plotone{f1.eps}~\\\bigskip\centerline{Fig. \ref{f:profile}}
\newpage
\plotone{f2.eps}~\\ \bigskip\centerline{Fig.\ref{f:j-vs-s}}
\newpage
\plotone{f3.eps}~\\ \bigskip\centerline{Fig. \ref{f:j-vs-mdot}}
\newpage
\plotone{f4.eps}~\\ \bigskip\centerline{Fig.\ref{f:lumin}}
\newpage
\plotone{f5.eps}~\\\bigskip\centerline{Fig. \ref{f:profiles2}}

\end{document}